\documentclass[sigconf]{acmart}

\def\BibTeX{{\rm B\kern-.05em{\sc i\kern-.025em b}\kern-.08emT\kern-.1667em\lower.7ex\hbox{E}\kern-.125emX}}
    
\acmYear{} 
\acmPrice{} 
\acmISBN{} 
\acmDOI{} 
\setcopyright{none} 
\acmConference[AdvML'19: Workshop on Adversarial Learning Methods for Machine Learning and Data Mining at KDD]{AdvML'19: Workshop on Adversarial Learning Methods for Machine Learning and Data Mining at KDD}{August 5th, 2019}{Anchorage, Alaska, USA}

\usepackage{amsmath}
\usepackage{mathrsfs}

\DeclareMathAlphabet\mathbfcal{OMS}{cmsy}{b}{n}

\begin{document}

%
\title{Defending against Backdoor Attack on Deep Neural Networks
}

%


 

\author{Hao Cheng$^{1}$, Kaidi Xu$^{2}$, Sijia Liu$^{3}$, Pin-Yu Chen$^{3}$, Pu Zhao$^{2}$, Xue Lin$^{2}$}
\affiliation{
    \institution{
    $^{2}$Xi'an Jiaotong University, Xi'an, China\\
    $^{2}$Northeastern University, Boston, US\\
    $^{3}$MIT-IBM Watson AI Lab, Cambridge, US
    }
}
\email{hectorcheng@stu.xjtu.edu.cn, {xu.kaid, zhao.pu}@husky.neu.edu, {sijia.liu, pin-yu.chen}@ibm.com, xue.lin@northeastern.edu}







%
\renewcommand{\shortauthors}{Hao Cheng, Kaidi Xu, Sijia Liu, et al.}

%
\begin{abstract}
Although deep neural networks (DNNs) have achieved a great success in various computer vision tasks, it is recently found that they are vulnerable to adversarial attacks.
In this paper, we focus on the so-called \textit{backdoor attack}, which injects a backdoor trigger to a small portion of training data (also known as data poisoning) such that the trained DNN induces misclassification while facing examples with this trigger.  
To be specific, we carefully study the effect of both real and synthetic backdoor attacks on the internal response of vanilla and backdoored DNNs through the lens of Gard-CAM.
Moreover, we show that the backdoor attack induces a significant bias in neuron activation in terms of the $\ell_\infty$
norm of an activation map compared to its $\ell_1$ and $\ell_2$ norm. Spurred by our results, we propose  the  \textit{$\ell_\infty$-based neuron pruning} to remove the backdoor from the backdoored DNN.
Experiments show that our method could effectively decrease the attack success rate, and also hold a high classification accuracy for clean images. 
\end{abstract}

%
%




%

%

%
\maketitle

\section{Introduction}

{Deep learning or deep neural network (DNN), as an outstanding machine learning technique, has become a foundational means for solving grand societal challenges, revolutionizing many application domains with superior performance~\cite{ding2017circnn,shi2018pairwise,ding2019req,sun2019hsim}.
Just like for traditional machine learning techniques, the security for deep learning is of great importance to its broad deployments, especially in the security-critical domains.
Since 2014, when Szegedy et al. \cite{szegedy2014intriguing} and subsequent work \cite{goodfellow2015explaining,nguyen2015deep,xu2018structured} made the discovery of adversarial examples against DNNs, an ever-increasing amount of research effort has been devoted to the design and countermeasures of the so-called DNN evasion (adversarial) attacks~\cite{xu2019topology,xu2019interpreting,ye2019adversarial,zhao2019admm,zhao2019design}.
}

{Another important category of adversarial attacks against DNNs is the data poisoning (adversarial) attack \cite{xiao2015feature, munoz2017towards, shafahi2018poison},  which results in illy-trained DNNs from the poisoned training dataset.
The \emph{backdoor attack} is a special type of data poisoning attack with better stealthiness and attacker controllability \cite{gu2017badnets}.
The backdoor attack is implemented through both pre-training and post-training processes.
In the pre-training process, poisoned training data is prepared by patching clean images with a particular trigger pattern and labelling such images with the trigger as a target wrong label.
Such prepared poisoned training data will be added into the training dataset without the awareness of the dataset users, and therefore DNNs trained from this poisoned training dataset become the backdoored DNNs.
In the post-training process, a backdoored DNN when presented with an image with the trigger will predict it into the target wrong label even if the trigger has a small size. It is expected that a backdoored DNN predicts clean images like a vanilla DNN, without noticeable mis-behaviors.
}

{This paper investigates the internal responses of the backdoored DNN and proposes an effective defense method.
We start from characterizing the vanilla and backdoored DNNs through the Grad-CAM  \cite{selvaraju2017grad} using different input and label combinations. The triggers are synthesized using the trigger reverse engineering method in \cite{wang2019neural}.
We found visually that the discriminative area of the backdoored DNN will be on the trigger region, indicating a higher activation value of some neurons within the network.
The visual and qualitative results from Grad-CAM inspire us for further quantitative analysis.
Then we plot the neuron activation map of the backdoored DNN using clean images with and without the trigger and analyze the $\ell_p$ norm of neuron activation values statistically.
And we found that the $\ell_\infty$ norm demonstrates the most significant difference between clean images and images with the trigger.
Therefore, the $\ell_\infty$ based neuron pruning is proposed as a defense against the backdoor attack.
We find the optimal pruning threshold value for the trade-off between the test accuracy on clean images and the attack success rate.
We can decrease the attack success rate from 81.6$\%$ to 48.42$\%$ with minor accuracy loss for the clean images. 
}

{The contributions of this work are summarized as follows: (i) We leverage Grad-CAM to visualize the relationship between images with and without trigger with respect to true and target labels on the vanilla DNN and the backdoored DNN. (ii) Further quantitative analysis based on neuron activation values demonstrates the $\ell_\infty$ norm is the best criteria for neuron pruning as a defense. (iii) We significantly reduce the attack success rate by the $\ell_\infty$ based neuron pruning.
}

\section{Backdoor Attack}

{In this section, we review the related work on the backdoor attack and also propose the threat model for this work.}

\subsection{Related Work}

{The initial backdoor attack was first proposed by Gu, Dolan-Gavitt, and Garg \cite{gu2017badnets}, which uses a pre-defined trigger pattern, such as a sticker on the traffic signs. 
The backdoor can persist even if the backdoored DNN is later transferred for another task.
Liu, Ma, Aafer, et al. demonstrated how to obtain a backdoored DNN from a vanilla DNN without tampering with the original training process \cite{liu2017trojaning}.
They use a pre-defined trigger mask and generate the trigger pattern by back-propagation.
Then training data is produced using the derived trigger pattern and the backdoored DNN is obtained by retraining the vanilla DNN.
} 

{Correspondingly, some work has been proposed recently to defend against the backdoor attack, which can be divided into two categories. The first category is to examine the untrusted training dataset through analyzing spectral signatures \cite{tran2018spectral} and activation clustering \cite{chen2018detecting}. 
The second category of work aims to modify a backdoored DNN to remove the backdoor such as
neural cleanse \cite{wang2019neural} and fine-pruning \cite{liu2018fine}. 
}

{Different from the previous work  \cite{wang2019neural,liu2018fine,tran2018spectral,chen2018detecting}, our paper places a significant emphasis on analyzing and explaining the effects of backdoor attack (original or synthetic) on both vanilla and backdoored DNNs. 
We also revisit the idea of neuron activation pruning and find that the $\ell_\infty$-norm based neuron pruning is the most effective one compared to $\ell_1$ and $\ell_2$ based scheme.
}

\subsection{Threat model}\label{sec:threatmodel}

{In this work, we target at the removal of backdoor from the backdoored DNN as a defense against backdoor attack.
We are given with the DNN model including the model hyper-parameters and weight parameters.
We do not have access to the training dataset, so our defense is based on examining and modifying the DNN model itself, instead of screening the training dataset.
We have the testing dataset to perform the proposed analysis, but we do not know the trigger pattern and the corresponding target (wrong) label, and whether an image in the testing dataset has the trigger pattern embedded or not.
}

\begin{figure}[tb]
\begin{minipage}[b]{0.2\linewidth}
  \centering
  \centerline{\includegraphics[width=1.8cm]{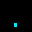}}
  \centerline{(a)}\medskip
\end{minipage}
\hfill
\begin{minipage}[b]{0.2\linewidth}
  \centering
  \centerline{\includegraphics[width=1.8cm]{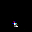}}
  \centerline{(b)}\medskip
\end{minipage}
\hfill
\begin{minipage}[b]{.2\linewidth}
  \centering
  \centerline{\includegraphics[width=1.8cm]{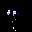}}
  \centerline{(c)}\medskip
\end{minipage}
\hfill
\begin{minipage}[b]{0.2\linewidth}
  \centering
  \centerline{\includegraphics[width=1.8cm]{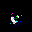}}
  \centerline{(d)}\medskip
\end{minipage}

\caption{Original and synthetic triggers: (a) original trigger for the target label 8; (b) synthetic trigger for the target label 8; (c) synthetic trigger for the label 14 (not the target label); and (d) synthetic trigger for the label 38 (not the target label).
}
\label{fig: fig1_example_backdoor}
\end{figure}

{Because we have no information about the trigger pattern, we employ the reverse-engineering method of the trigger pattern in \cite{wang2019neural} to synthesize the trigger pattern.
Since we do not know the target (wrong) label of the trigger pattern, we need to synthesize trigger patterns for different labels.
Figure \ref{fig: fig1_example_backdoor} demonstrates the original trigger and some synthetic triggers.
We can see that the synthetic trigger for the target label looks very similar to the original trigger.
We may calculate the $\ell_1$ norms of the synthetic triggers to determine the target label.
}


{Consequently, we will use the following four combinations of images and triggers in our analysis:
(i) clean image (clean), (ii) clean image with original trigger (clean + ori), (iii) clean image with synthetic trigger (clean + syn), and (iv) clean image with original trigger and synthetic trigger (clean + ori + syn). 
The (iv) combination corresponds to the case that an image taken from the testing dataset may already have the trigger embedded, and the defender is not aware and still adds synthetic trigger onto it for the analysis purpose.
}


\section{Grad-CAM Analysis}
{We use Grad-CAM \cite{selvaraju2017grad} to visually demonstrate the DNN's discriminative area.
Compared with the original Class Activation Mapping (CAM) \cite{zhou2015cnnlocalization}, Grad-CAM has a better applicability for complicated DNN architectures and therefore is chosen for our analysis.
Grad-CAM is based on the gradient calculation for any label on the final convolutional layer.
}

\begin{figure*}[htb]
\centering

\begin{minipage}[b]{0.1\linewidth}
  \centering
  \centerline{[clean,}
  \centerline{true]}
  \centerline{\includegraphics[width=2cm]{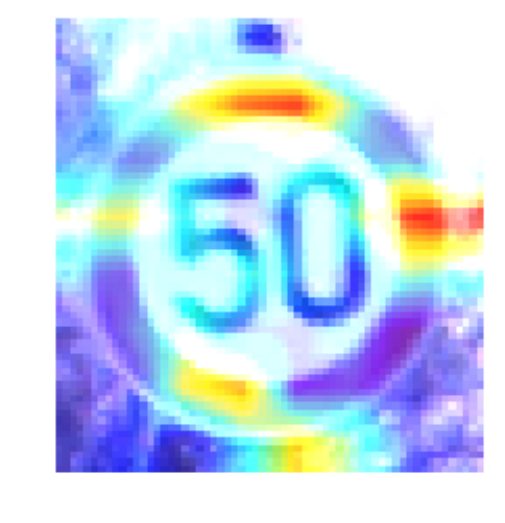}}
  \centerline{(a)}\medskip
\end{minipage}
\hfill
\begin{minipage}[b]{0.1\linewidth}
  \centering
  \centerline{[clean,}
  \centerline{target]}
  \centerline{\includegraphics[width=2cm]{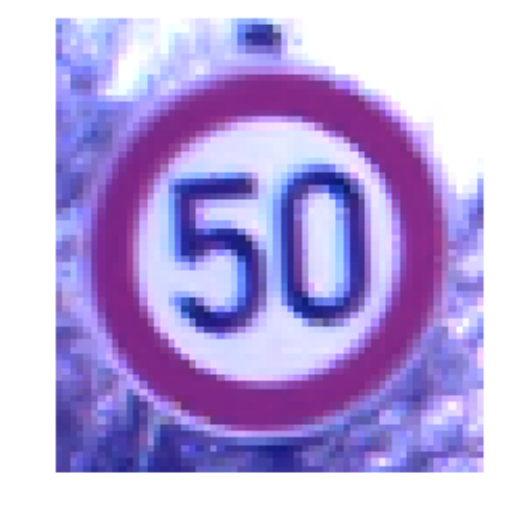}}
  \centerline{(b)}\medskip
\end{minipage}
\hfill
\begin{minipage}[b]{0.1\linewidth}
  \centering
  \centerline{[clean+ori,}
  \centerline{true]}
  \centerline{\includegraphics[width=2cm]{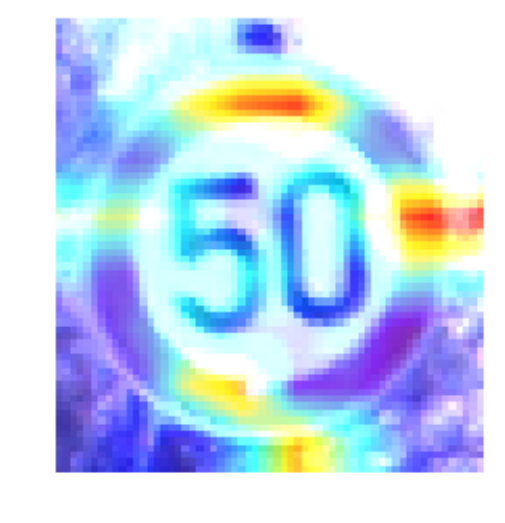}}
  \centerline{(c)}\medskip
\end{minipage}
\hfill
\begin{minipage}[b]{0.1\linewidth}
  \centering
  \centerline{[clean+ori,}
  \centerline{target]}
  \centerline{\includegraphics[width=2cm]{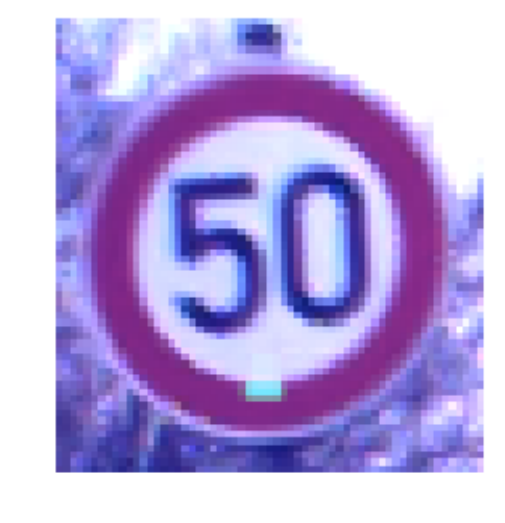}}
  \centerline{(d)}\medskip
\end{minipage}
\hfill
\begin{minipage}[b]{.1\linewidth}
  \centering
  \centerline{[clean+syn,}
  \centerline{ true]}
  \centerline{\includegraphics[width=2cm]{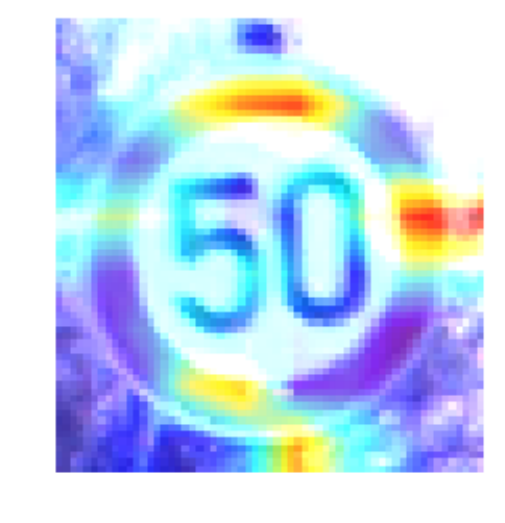}}
  \centerline{(e)}\medskip
\end{minipage}
\hfill
\begin{minipage}[b]{0.1\linewidth}
  \centering
  \centerline{[clean+syn,}
  \centerline{target]}
  \centerline{\includegraphics[width=2cm]{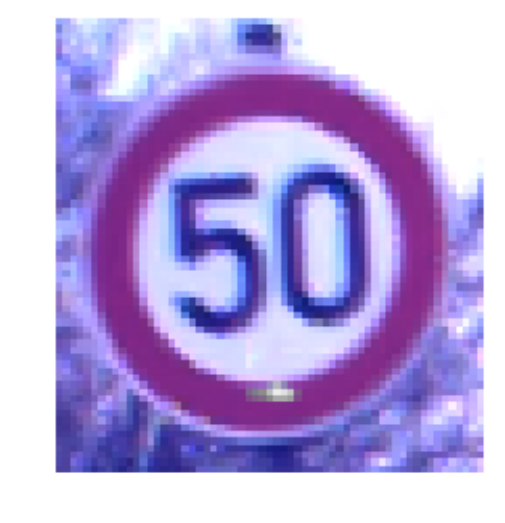}}
  \centerline{(f)}\medskip
\end{minipage}
\hfill
\begin{minipage}[b]{0.1\linewidth}
  \centering
  \centerline{[clean+ori+syn,}
  \centerline{ true]}
  \centerline{\includegraphics[width=2cm]{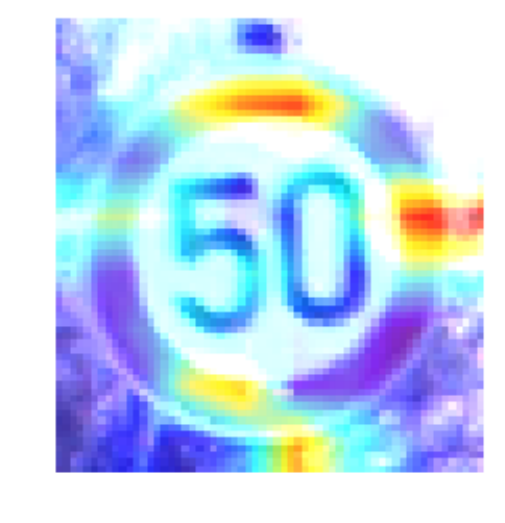}}
  \centerline{(g)}\medskip
\end{minipage}
\hfill
\begin{minipage}[b]{0.1\linewidth}
  \centering
  \centerline{[clean+ori+syn,}
  \centerline{ target]}
  \centerline{\includegraphics[width=2cm]{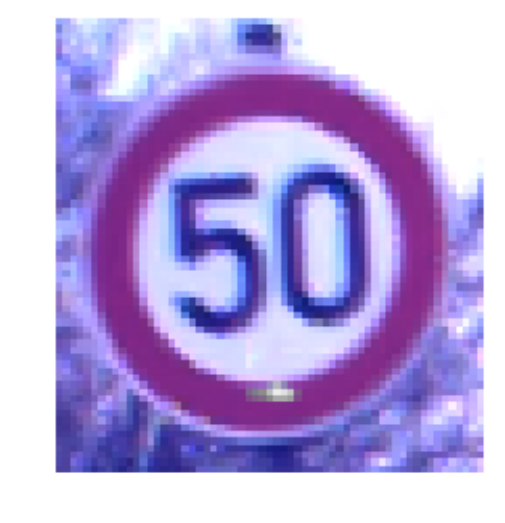}}
  \centerline{(h)}\medskip
\end{minipage}

\begin{minipage}[b]{0.1\linewidth}
  \centering
  \centerline{\includegraphics[width=2cm]{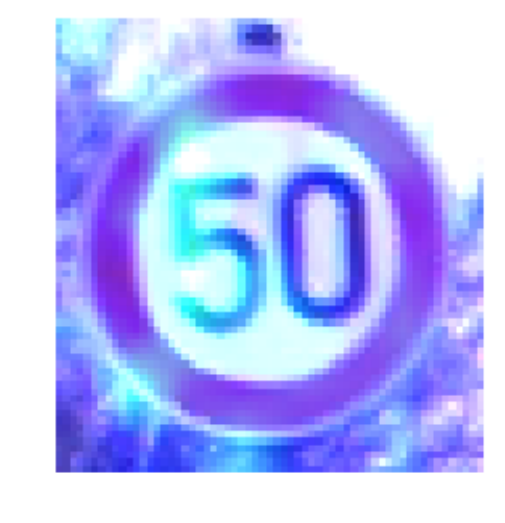}}
  \centerline{(a')}\medskip
\end{minipage}
\hfill
\begin{minipage}[b]{0.1\linewidth}
  \centering
  \centerline{\includegraphics[width=2cm]{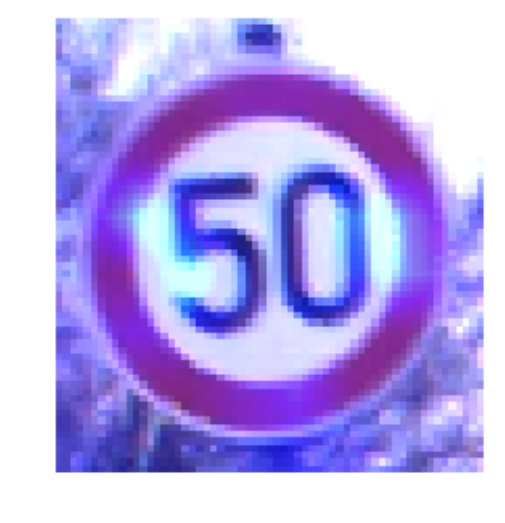}}
  \centerline{(b')}\medskip
\end{minipage}
\hfill
\begin{minipage}[b]{0.1\linewidth}
  \centering
  \centerline{\includegraphics[width=2cm]{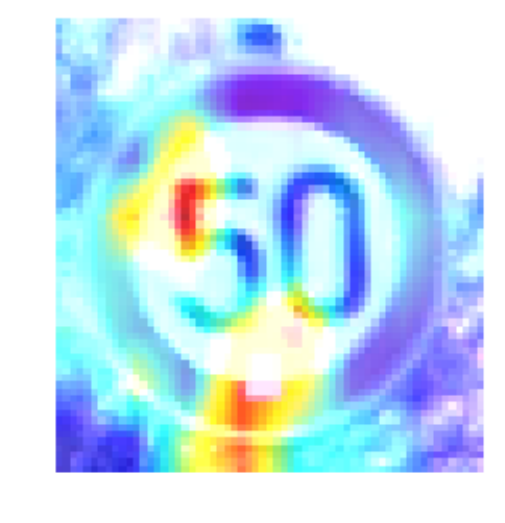}}
  \centerline{(c')}\medskip
\end{minipage}
\hfill
\begin{minipage}[b]{0.1\linewidth}
  \centering
  \centerline{\includegraphics[width=2cm]{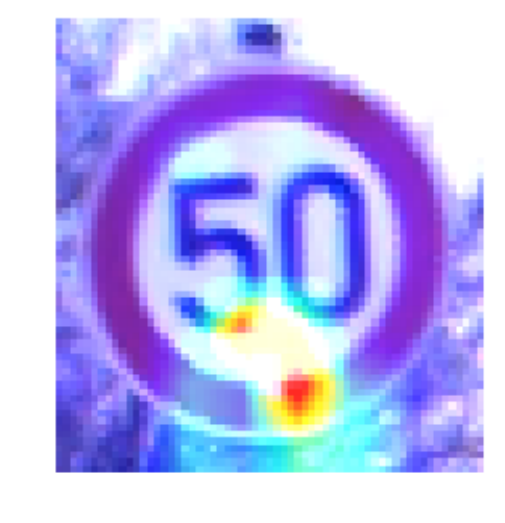}}
  \centerline{(d')}\medskip
\end{minipage}
\hfill
\begin{minipage}[b]{.1\linewidth}
  \centering
  \centerline{\includegraphics[width=2cm]{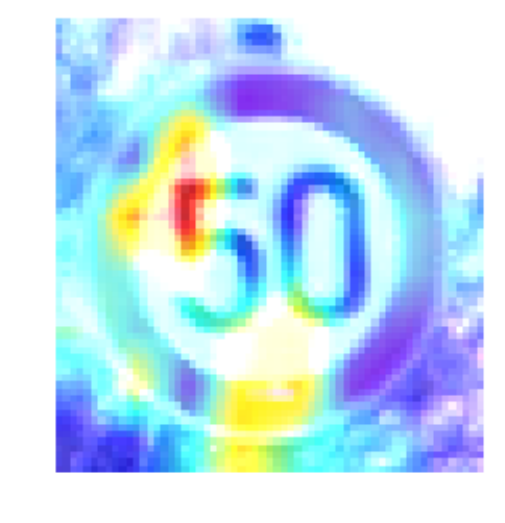}}
  \centerline{(e')}\medskip
\end{minipage}
\hfill
\begin{minipage}[b]{0.1\linewidth}
  \centering
  \centerline{\includegraphics[width=2cm]{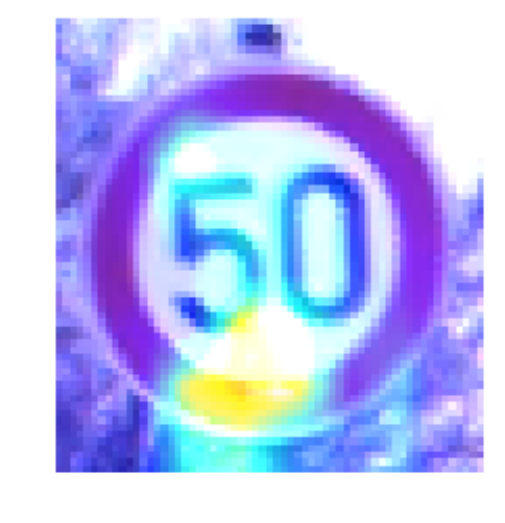}}
  \centerline{(f')}\medskip
\end{minipage}
\hfill
\begin{minipage}[b]{0.1\linewidth}
  \centering
  \centerline{\includegraphics[width=2cm]{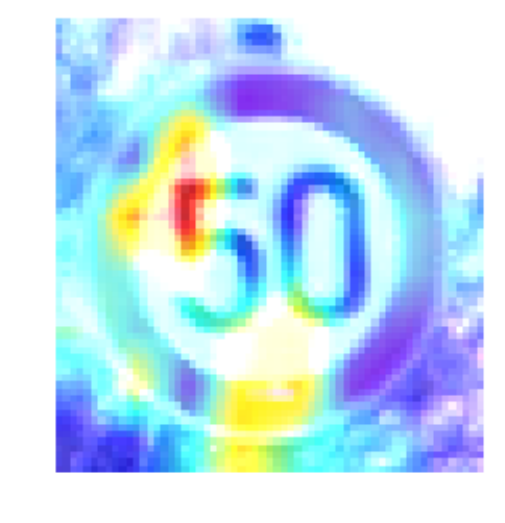}}
  \centerline{(g')}\medskip
\end{minipage}
\hfill
\begin{minipage}[b]{0.1\linewidth}
  \centering
  \centerline{\includegraphics[width=2cm]{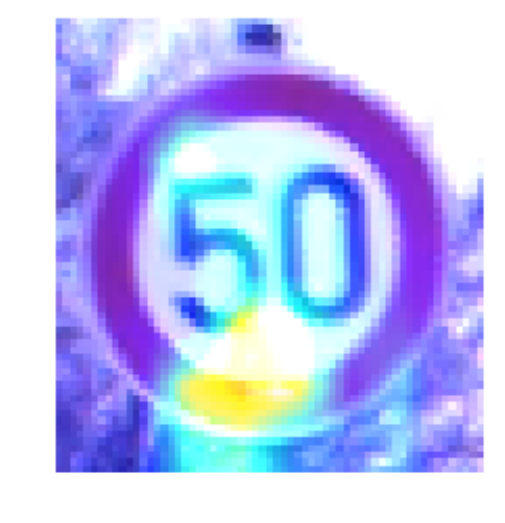}}
  \centerline{(h')}\medskip
\end{minipage}
\caption{Grad-CAM overlaid on top of the input images to DNN.
The first row (a)$\sim$(h) is from the vanilla DNN and the second row (a')$\sim$(h') is from the backdoored DNN.
On top of each column, the setting of (input, label) pair is noted. 
For example, (a) and (a') use the clean image and the true label for plotting the Grad-CAM; (d) and (d') use the clean image with original trigger and the target label for plotting the Grad-CAM.
}
\label{fig: fig4_gradcam}
%
\end{figure*}

{Figure \ref{fig: fig4_gradcam} shows the Grad-CAM overlaid on top of the input images. We use two DNN models: the vanilla DNN is used for plotting the first row of subfigures (a)$\sim$(h), and the backdoored DNN is used for plotting the second row of subfigures (a')$\sim$(h'). We use the four input settings discussed in Section \ref{sec:threatmodel}. For each of them, we use both the true label and the target label.
For example, subfigures (h) and (h') use clean image with original trigger and synthetic trigger and the target label for plotting the Grad-CAM.
}

{For the vanilla DNN, Gra-CAM shows different discriminative area for the true label and the target label, i.e., when we compare (a) with (b), (c) with (d), etc.
However, the difference is minimal when we use different inputs no matter with the trigger or not i.e., comparing (a), (c), (e), and (g).
And the vanilla DNN only responds to the true label of the input.
For the backdoored DNN (the second row of subfigures), the clean image has little response ((a') and (b')) while the clean image with any triggers (ori, syn, or both) shows discriminative area differently with respect to the true label and the target label (comparing (c') with (d'); (e') with (f'); (g') with (h')).
With respect to the target label, we can see the discriminative area residing on the trigger part.
}



\section{Neuron Activation Analysis: $\ell_\infty$ outlier}

{The visual and qualitative results from Grad-CAM inspire us for quantitative analysis.
For this purpose, we plot the activation map and characterize the $\ell_p$ norm of the activation values.
}



\begin{figure}[htb]

\begin{minipage}[b]{.52\linewidth}
  \centering
  \centerline{\includegraphics[width=8cm]{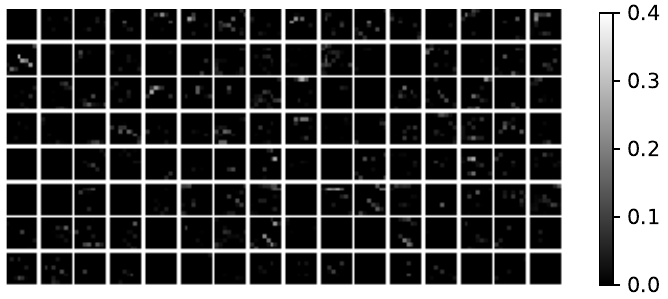}}
  \centerline{(a)}\medskip
\end{minipage}
\hfill
\begin{minipage}[b]{.52\linewidth}
  \centering
  \centerline{\includegraphics[width=8cm]{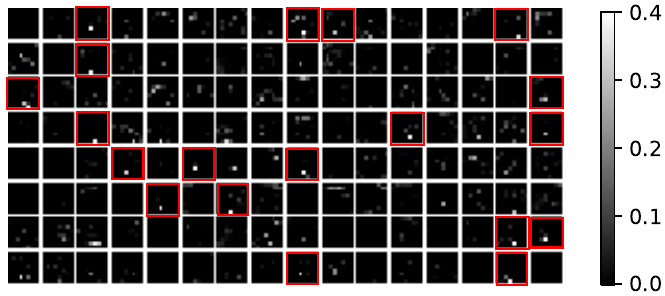}}
  \centerline{(b)}\medskip
\end{minipage}
\caption{Neuron activation map of the backdoored DNN using (a) clean image and (b) clean image with original trigger, for all the 128 neurons in the final convolutional layer.}
\label{fig: fig2_activation_map}
\end{figure}

{First, we plot the neuron activation map of the backdoored DNN using both clean image and clean image with the original trigger in Figure \ref{fig: fig2_activation_map}, each grid representing one neuron activation.
We can observe that some neurons demonstrate obvious activation in response to the trigger, and this fact further motivates us for quantitative analysis using $\ell_p$ norms.
}

\begin{figure*}[htb]

\begin{minipage}[b]{.32\linewidth}
  \centering
  \centerline{\includegraphics[width=6.7cm]{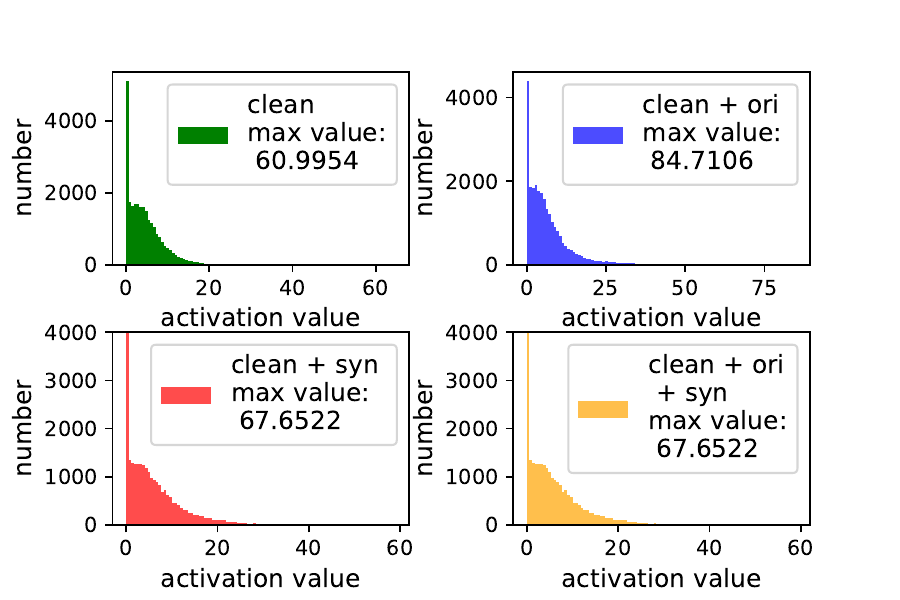}}
  \centerline{$\ell_1$}\medskip
\end{minipage}
\hfill
\begin{minipage}[b]{0.30\linewidth}
  \centering
  \centerline{\includegraphics[width=6.7cm]{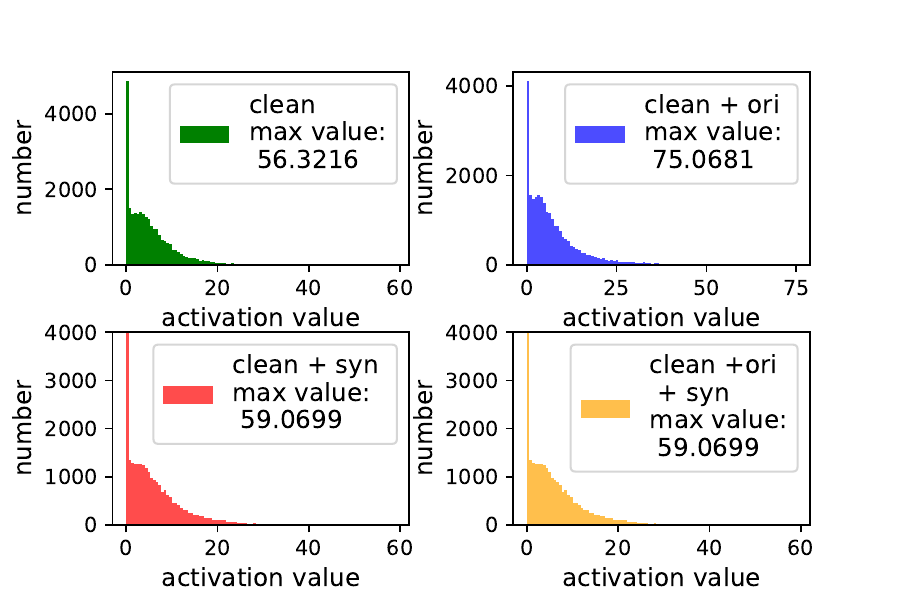}}
  \centerline{$\ell_2$}\medskip
\end{minipage}
\hfill
\begin{minipage}[b]{0.30\linewidth}
  \centering
  \centerline{\includegraphics[width=6.7cm]{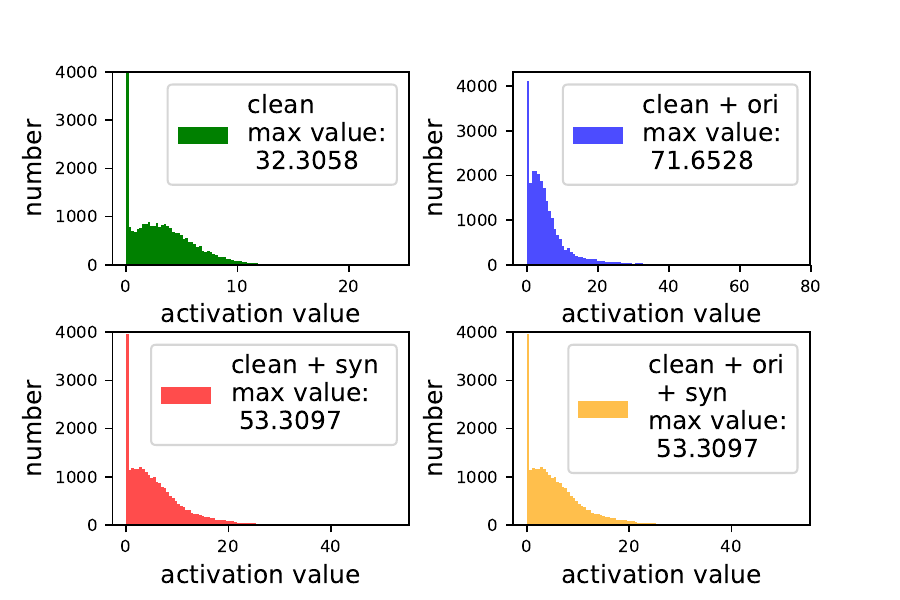}}
  \centerline{$\ell_\infty$}\medskip
\end{minipage}
\caption{Histogram of the $\ell_1$, $\ell_2$ and $\ell_\infty$ norms of the final convolutional layer activation values. Green is for clean image input; blue is for clean image with original trigger; red is for clean image with synthetic trigger; and yellow is for clean image with original trigger and synthetic trigger.
}
\label{fig: fig3_activation_ellinf}
\end{figure*}


{Figure \ref{fig: fig3_activation_ellinf} plots the histogram of the $\ell_1$, $\ell_2$ and $\ell_\infty$ norms of the final convolutional layer activation values. For each $\ell_p$ norm, the four input settings discussed in Section \ref{sec:threatmodel} are used for plotting the four histograms, one color for each input setting.
The following observations are made:
(i) For any $\ell_p$ norm, the maximum activation value (labelled with each histogram) is increased when trigger is added (no matter it is original trigger, synthetic trigger or both).
(ii) The increase is most significant in the $\ell_\infty$ norm case.
}

\section{$\ell_\infty$ based Neuron Pruning}

\subsection{Methodology}

\begin{figure}[htb!]

\begin{minipage}[b]{.52\linewidth}
  \centering
  \centerline{\includegraphics[width=7.2cm]{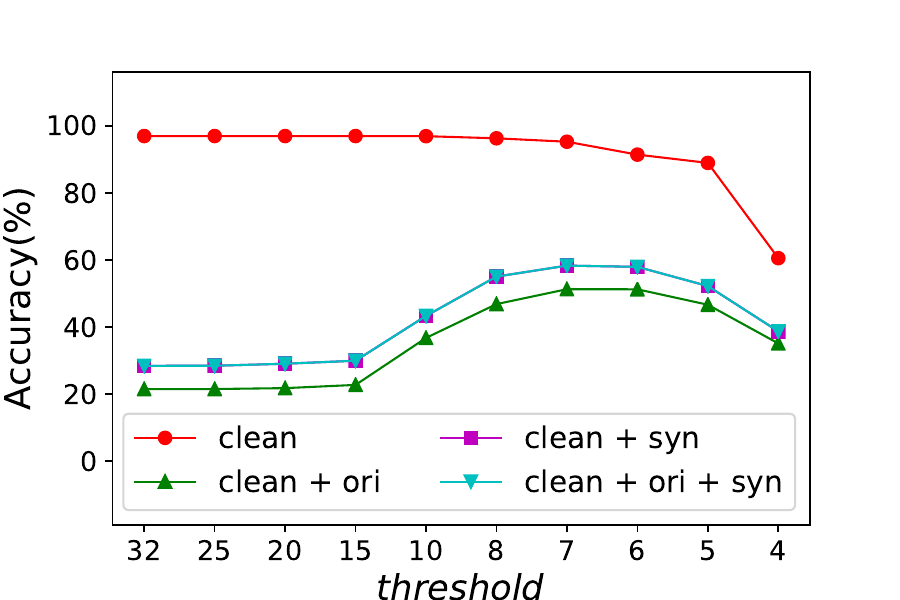}}
  \centerline{(a)}\medskip
\end{minipage}
\hfill
\begin{minipage}[b]{0.52\linewidth}
  \centering
  \centerline{\includegraphics[width=7.2cm]{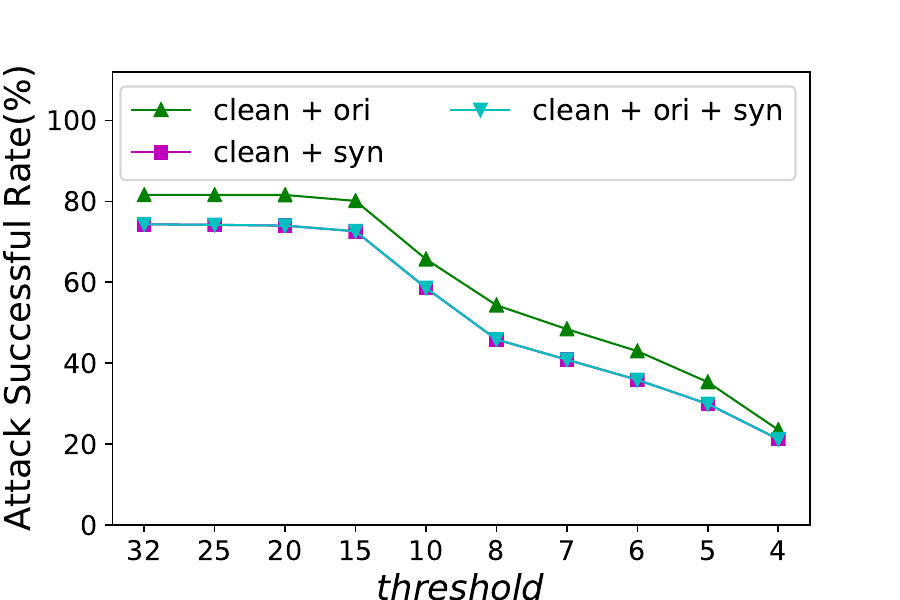}}
  \centerline{(b)}\medskip
\end{minipage}
\caption{(a) Classification accuracy for four input settings; and (b) attack successful rate for three input settings vs pruning threshold.}
\label{fig: fig5_pruning_results}
\end{figure}

{Based on previous observation that images with triggers will result in significant increase of the $\ell_\infty$ norm of the final convolutional layer activation values, we proposed to perform $\ell_\infty$ based neuron pruning to defend against backdoor attack.
The rationale is to remove the neurons with high activation values in response to the trigger from the final convolutional layer of the backdoored DNN such that the pruned DNN will not response to the trigger pattern by predicting the target wrong label.
The difficulty lies in selecting the pruning threshold of the $\ell_\infty$ norm of the neuron activation values.
In actual operation, we choose the initial threshold as the max value of clean images' activation value, $32.305782$, and gradually lower the threshold value to increase the defense effect while maintaining high classification accuracy of the clean images.
}


\subsection{Experimental Setting}
{In this paper, we focus on the traffic sign classification task. We use German Traffic Sign Recognition Benchmark (GTSRB) dataset. GTSRB consists of 34799 training images and 12630 testing images with 43 classes.
We select the AlexNet as our DNN model architecture.
The backdoored AlexNet is trained using the method in \cite{gu2017badnets} and a small square as the trigger pattern.
}

\subsection{Experimental Results}

{In Figure \ref{fig: fig5_pruning_results}, we present (a) accuracy and (b) attack success rate with respect to the pruning threshold.
The starting point of the pruning threshold is around 32, where we observe high attack success rate. 
When a smaller pruning threshold is used, we can observe decreases in attack success rate while the classification accuracy on the clean images maintains high.
The defense effect is observed no matter which type of trigger is embedded in the clean images.
From the figure, we find that at pruning threshold values of 6 and 7, we achieve the best trade-off between attack success rate and accuracy on clean images.
Furthermore, we summarize in Table \ref{table_mnist_lottery} the test accuracy and attack success rate of a backdoored DNN and a backdoored-and-pruned DNN.
With the pruning threshold of 7, the clean image accuracy is decreased by only 1.7$\%$ while the attack success rate is decreased from 81.61$\%$ to 48.42$\%$.
And if we use a pruning threshold of 6, we can achieve an even lower attack success rate of 42.99$\%$ but with the penalty of more testing accuracy loss.
}

\begin{table}[htb]
\small
 \caption{The test accuracy of clean images and the attack success rate (SR) in $\%$ with and without the $\ell_\infty$ based neuron pruning.
  } 
  \label{table_mnist_lottery}
 \centering
\begin{tabular}{c|cccc}
\toprule[1pt]
Threshold    & acc   & SR(clean+ori)  & SR(clean+syn)          & SR(clean+ori+syn)           \\
\midrule[1pt]

None   & 96.91 & 81.61  & 74.36   & 74.36           \\
7   & 95.21 & 48.42 & 40.87 & 40.87\\
6   & 91.38 & 42.99 & 35.90          & 35.90           \\
\bottomrule[1pt]
\end{tabular}
\end{table}

\section{Conclusion And Future Work}
This paper investigates the internal responses of the backdoored DNN and proposes an effective defensive method.
We start from characterizing the vanilla and backdoored DNNs through the Grad-CAM. 
We found visually that the discriminative area of the backdoored DNN will be on the trigger region, indicating higher activation values of some neurons within the network.
Then we plot the neuron activation map of the backdoored DNN using clean images with and without the trigger and analyze the $\ell_p$ norm of neuron activation values statistically.
And we found that the $\ell_\infty$ norm demonstrates the most significant difference between clean images and images with the trigger.
Therefore, the $\ell_\infty$ based neuron pruning is proposed as a defense against the backdoor attack.
We find the optimal pruning threshold value for the trade-off between the test accuracy on clean images and the attack success rate.

Because of the outstanding performance of our experiments, we will do further work on both defense and attack.  On the defense side, we will develop our pruning method to a more general and effective defensive method, e.g. developing a kind of robust training measure that refers to the $\ell_\infty$ gap of activation value between vanilla and backdoored DNNs. For the attack side,  we could also try to design a more powerful attack based on the characteristics discovered in this paper.


%

%
\bibliographystyle{ACM-Reference-Format}
\bibliography{reference}

\end{document}